\documentstyle[aps,prl,epsfig,twocolumn]{revtex}

\begin{document}
\draft
\wideabs{
\title{ Critical Behavior of a Heavy Particle in a Granular Fluid}
\author{Andr\'{e}s Santos\cite{andres} and James W. Dufty\cite{jim}}
\address{Department of Physics, University of Florida, Gainesville, FL 32611}
\date{\today}
\maketitle

\begin{abstract}
Behavior analogous to a
second order phase transition is observed for the homogeneous cooling state
of a heavy impurity particle in a granular fluid. The order parameter $\phi $
is the ratio of impurity mean square velocity to that of the fluid, with a
conjugate field $h$ proportional to the mass ratio. A parameter $\beta $,
measuring the fluid cooling rate relative to the impurity--fluid collision
rate, is the analogue of the inverse temperature. For $\beta <1$ the fluid
is ``normal'' with $\phi =0$ at $h=0$, as in the case of a system with
elastic collisions. For $\beta >1$ an ``ordered'' state with $\phi \neq 0$
occurs at $h=0$, representing an extreme breakdown of equipartition.
Critical slowing and qualitative changes in the velocity distribution
function for the impurity particle near the transition are noted.
\end{abstract}

}
The simplest statistical mechanical model for an activated granular fluid is
a system of smooth, inelastic hard spheres. The inelasticity is specified in
terms of a restitution coefficient $\alpha \leq 1$. Qualitative differences
between the cases $\alpha =1$ (elastic spheres) and $\alpha \neq 1$ have
been demonstrated in many ways using both theoretical and simulation
methods \cite{general}. Among these is the replacement of the Gibbs state for an isolated
uniform system by a homogeneous cooling state (HCS). The HCS is inherently
time dependent due to the loss of energy in each binary collision. It is
postulated that the time dependence of this ensemble occurs entirely
through a scaling of the velocity by its root mean square (``thermal'')
velocity $v_{T}(t)$ and the associated normalization \cite{Brey1}. In this
case, the dynamics (Liouville equation or kinetic equation) can be
transformed to a stationary state form using suitable dimensionless
variables. The time dependence of $v_{T}(t)$ also follows from
this scaling, so that only a time independent dimensionless cooling rate
must be determined self-consistently.  Similar results apply for a multi-component system, although the
analysis is complicated by the existence of \ many scaling velocities \cite
{Garzo1}. The HCS results when the cooling rates for all 
mean square velocities are the same. This condition then determines all
scaling velocities in terms of that for one species. It is the analogue of
energy
equipartition for a system with $\alpha =1$, but deviations from
equipartition occur for all $\alpha \neq 1$. These deviations can be weak or
strong depending on the mechanical differences between the species and the
degree of dissipation.

\begin{figure}[t]
\centerline{\epsfig{file=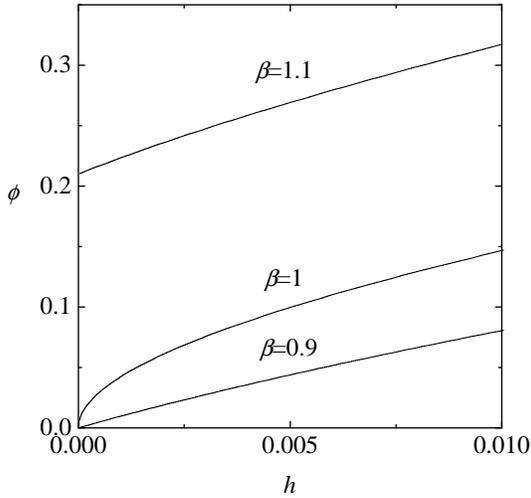,width=7.0cm}}
\caption{Ratio of mean square velocities, $\phi$, as a function of the mass ratio parameter 
$h$ for $\beta=0.9$, 1, and 1.1.
\label{fig1}}
\end{figure}

Many of the new effects to be expected in multi-component systems can be
illustrated by the simplest case of a single impurity particle in a one
component system \cite{Brey3}. The fluid (impurity) particle diameter, mass, and
restitution coefficient are denoted by $\sigma $ ($\sigma _{0}$) , $m$ ($m_{0}
$), and $\alpha $ ($\alpha _{0}$). There are two mean square velocities and
associated cooling rates for the fluid and impurity particles, 
\begin{equation}
\begin{array}{l}
v_{T}^{2}(t) =\frac{2}{3}\langle v^{2}(t)\rangle,\quad 
v_{T0}^{2}(t)=\frac{2}{3}\langle v_{0}^{2}(t)\rangle ,  \\
\xi  =-\partial _{t}\ln v_{T}^{2}(t),\quad \xi _{0}=-\partial
_{t}\ln v_{T0}^{2}(t). 
\end{array}
\label{1}
\end{equation}
The brackets denote an average over the HCS, and the subscript $0$ denotes
properties of the impurity particle. The cooling rates for a moderately
dense fluid can be calculated to good accuracy from the Boltzmann-Enskog
and Lorentz-Enskog kinetic equations for the fluid and impurity reduced
distribution functions \cite{Garzo1}. 
These rates then depend on the velocity distributions for the fluid and impurity particle in
the HCS and can be estimated to good approximation using Maxwellians (see
justification below) with the results 
\begin{equation}
\xi ^{\ast }=\frac{\xi }{n\sigma ^{2}gv_{T}}=\frac{2\sqrt{2\pi }}{3}
\left( 1-\alpha ^{2}\right) ,\hspace{0.3in}  \label{2}
\end{equation}
\begin{equation}
\xi _{0}^{\ast }=\frac{\xi _{0}}{n\sigma ^{2}gv_{T}}=\nu^*\left(
1+\phi \right) ^{1/2}\left( 1-h\frac{1+\phi }{\phi }\right),   \label{3}
\end{equation}
where $n$ is the fluid density, $g$  is the pair correlation
function for two fluid particles,
and $n\sigma ^{2}gv_{T}$ is a characteristic collision rate
for the fluid particles. Similarly, $\nu^{\ast }=\nu /n\sigma ^{2}gv_{T}=(16
\sqrt{\pi }/3)h\left(g_{0}/g\right) \left(\overline{\sigma }/\sigma \right)
^{2}$ is a corresponding dimensionless average impurity--fluid particle
collision rate, where $\overline{\sigma }=\left( \sigma +\sigma _{0}\right)
/2$ and $g_{0}$ is the corresponding fluid--impurity particle pair
correlation function.  The dependence on
the mass ratio occurs entirely through $h=m(1+\alpha _{0})/2(m+m_{0})$. The
time independent ratio of mean square velocities, 
\begin{equation}
\phi =v_{T0}^{2}/v_{T}^{2},  \label{4}
\end{equation}
is determined by the condition that the cooling rates be equal, 
\begin{equation}
h=\frac{\phi }{\left( 1+\phi \right) ^{3/2}}\left[ \left( 1+\phi \right)
^{1/2}-\beta \right].   \label{5}
\end{equation}
The constant $\beta $ is a measure of the fluid cooling rate  relative to the
fluid--impurity collision rate,
\begin{equation}
\beta = \frac{\xi^*}{\nu^*}=\frac{\left( 1-\alpha ^{2}\right) }{
4\sqrt{2}h}\frac{g}{g_{0}}\left( \frac{\sigma }{\overline{\sigma }}\right)
^{2} .  \label{5a}
\end{equation}
 The elastic collisions limit is given by $\beta =0$, 
for which the solution
is $\phi =h/\left( 1-h\right) =m/m_{0}$. This is the required result from
the equipartition theorem. If only the fluid particle collisions are elastic
($\beta =0$, $\alpha _{0}\neq 1$), a recent exact result of Martin and Piasecki, $\phi
=m(1+\alpha _{0})/\left[ 2m_{0}+m\left( 1-\alpha _{0}\right) \right] $ is
recovered \cite{Martin}. More generally, Eq.\ (\ref{5}) determines $\phi
(\beta,h )$ for the entire physical domain of $h$ and $\beta $. 
The solution
is found to have qualitatively different behavior at small $h$ for different
values of $\beta $. The asymptotic behavior for $h\rightarrow 0$ is found to
be 
\begin{equation}
\phi (\beta,h )\rightarrow \left\{ 
\begin{array}{ll}
({1-\beta })^{-1}{h},&\beta <1 \\ 
\sqrt{2h},&\beta =1 \\ 
\beta ^{2}-1+2\beta^4(\beta^2-1)^{-1}{h},&\beta >1.
\end{array}
\right.   \label{6}
\end{equation}
Thus $\phi $ vanishes at $h=0$ for $\beta <1$ but is finite at $\beta >1$.
The condition $\beta <1$ is ``normal'' in the sense that $\phi $ scales as
the mass ratio just as for systems with elastic collisions and the
consequent equipartition of kinetic energy. This normal state has been
studied recently for $h\ll 1$, where the kinetic theory implies a
corresponding Fokker-Planck description \cite{Brey3}. The condition 
$\beta >1$ corresponds to an extreme violation of equipartition such that the
mean square velocities of the fluid and impurity particles remain comparable
even though their mass ratio is zero. Figure \ref{fig1} shows the more complete
dependence of $\phi (\beta,h )$ on $h$ for  $\beta =0.9$, 1, and $1.1$,
confirming the above asymptotic analysis.
\begin{figure}[tbh]
\centerline{\epsfig{file=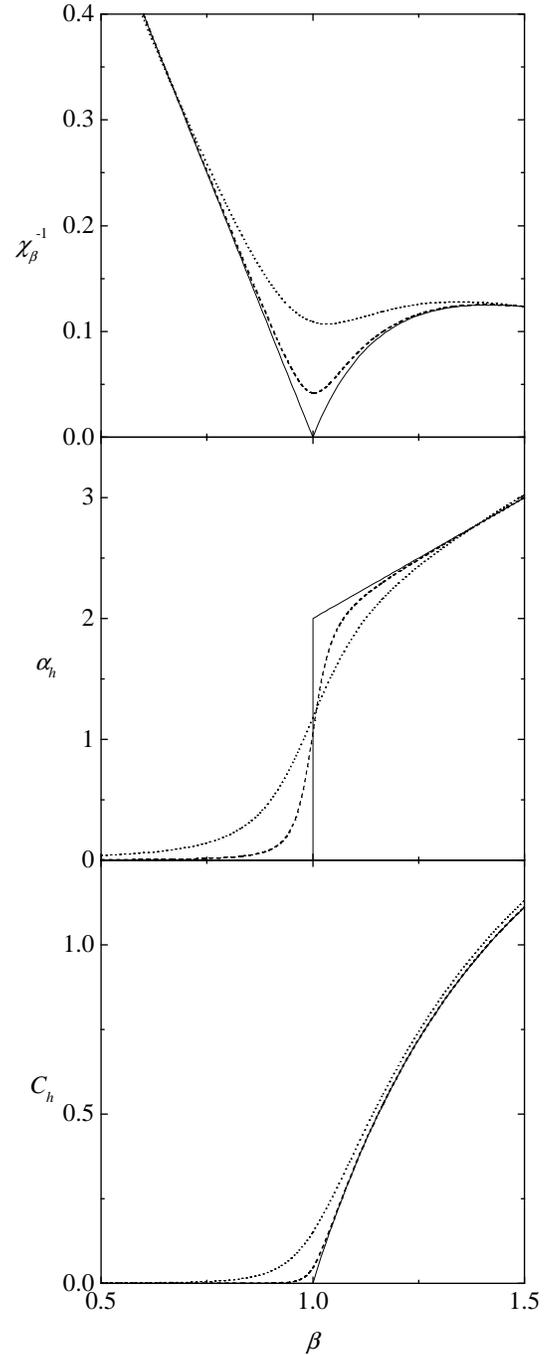,width=7.0cm}}
\caption{Inverse susceptibility ($\chi_\beta^{-1}$), expansion coefficient ($\alpha_h$), 
and heat capacity ($C_h$) as functions of $\beta$ for $h=10^{-2}$ (dotted line), 
$h=10^{-3}$ (dashed line), and $h=0$ (solid line).
\label{fig2}}
\end{figure}

This behavior at small $h$ is analogous to a thermodynamic transition at $%
\beta =1$ between two different states of the system characterized by the
order parameter $\phi $, conjugate field $h$, and inverse ``temperature'' $\beta 
$. To develop this analogy a Helmholtz free energy $F\left( \beta ,\phi
\right) $ is obtained from the ``equation of state'' (\ref{5}) by
integrating the definition $h=\partial F\left( \beta ,\phi
\right) /\partial \phi $  [with the boundary condition $F(\beta,0)=0$, 
for simplicity]. Next, the Gibbs free energy is constructed from $F$
by the Legendre transformation $\Phi \left( \beta ,h\right) =F-h\phi $. The
first and second derivatives of $\Phi \left( \beta ,h\right) $ provide the
order parameter $\phi $, ``entropy'' $\Sigma $, ``susceptibility'' $\chi _{\beta }$,
``expansion coefficient'' $\alpha _{h}$, and ``heat capacity'' $C_{h}$. The results
are 
\begin{equation}
\Phi \left( \beta ,h\right) =\left( 1-h\right) \phi -\ln \left(
1+\phi \right) -2\beta \left[\frac{2+\phi }{\left( 1+\phi \right) ^{1/2}}-2\right],
\label{9}
\end{equation}
\begin{equation}
\Sigma \left( \beta ,h\right) =\frac{\partial \Phi \left( \beta ,h\right) }{%
\partial \beta }=-2\left[\frac{2+\phi }{\left( 1+\phi \right) ^{1/2}}-2\right],
\label{10}
\end{equation}
\begin{equation}
\chi _{\beta }=-\frac{\partial ^{2}\Phi \left( \beta ,h\right) }{\partial
h^{2}}=\frac{\left( 1+\phi \right) ^{5/2}}{\left( 1+\phi
\right) ^{1/2}-\beta \left( 1-\frac{1}{2}\phi \right) },  \label{12}
\end{equation}
\begin{equation}
\alpha_h =-\frac{\partial ^{2}\Phi \left( \beta ,h\right) }{\partial \beta
\partial h}=\chi _{\beta }\frac{\phi }{\left( 1+\phi \right) ^{3/2}},
\label{13}
\end{equation}
\begin{equation}
C_{h}=-\frac{\partial ^{2}\Phi \left( \beta ,h\right) }{\partial \beta ^{2}}
=\chi _{\beta }^{-1}\alpha_h^2.
\label{14}
\end{equation}
The behavior of these functions near $h=0$ follows directly from (\ref{6}).
For example, the response functions $\chi _{\beta }$,  $\alpha
_{h}$, and $C_{h}$ at $h=0$, $10^{-2}$,  and $10^{-3}$ are shown in Fig.\ \ref{fig2}. 
In the limit $h\to 0$, the entropy and order
parameter are continuous at $\beta =1$, so the divergence of $\chi_\beta$ and
the discontinuity of $\alpha _{h}$
at $\beta =1$ identifies this as a second order phase transition.
 Near the critical region ($h\ll 1$, $|\beta-1|\ll 1$), the free energy adopts the 
Landau-like form $\Phi(\beta,h)\approx \frac{1}{2}(1-\beta)\phi^2+\frac{1}{6}\phi^3-h\phi$, 
which yields $h\approx (1-\beta)\phi+\frac{1}{2}\phi^2$. As a consequence,  
the free energy and the equation of state in the critical region satisfy the
scaling relations $\Phi(\lambda (\beta-1),\lambda^a h)=\lambda^b
\Phi(\beta-1,h)$ and 
$\phi(\lambda(\beta-1),\lambda^a h)=\lambda^{b-a} \phi(\beta-1,h)$, with $a=2$
and $b=3$. These scaling relations suffice to determine the critical exponents
\cite{note2} $\hat{\delta}=a/(b-a)=2$, $\hat{\beta}=b-a=1$, and 
$\hat{\gamma}=2a-b=1$, while the
critical exponent $\hat{\alpha}=2-b=-1$ is negative, indicating that $C_h$ is continuous
at the critical point.

\begin{figure}[t]
\centerline{\epsfig{file=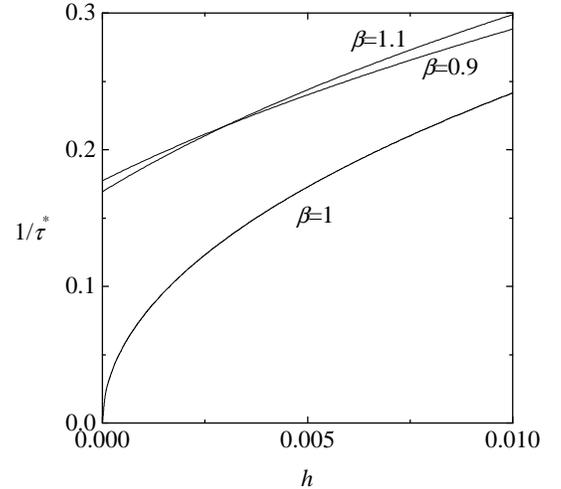,width=7.0cm}}
\caption{Inverse characteristic time ${\tau^*}^{-1}\equiv (\nu^*
\tau)^{-1}$ as a function of the mass ratio parameter 
$h$ for $\beta=0.9$, 1, and 1.1.
\label{fig3}}
\end{figure}

If the ratio between the initial mean square velocities of the fluid and impurity particles is
not that given by the solution to (\ref{5}), there is an evolution to the
HCS described by (\ref{1}) which can be written in the Ginzburg-Landau
form 
\begin{equation}
\partial _{s}\phi =-\left[ \xi _{0}^{\ast }(t)-\xi ^{\ast }(t)\right] \phi
=-\mu (\phi )\frac{\partial \Phi \left( \beta ,h;\phi \right) }{\partial
\phi }, \label{15}
\end{equation}
where a new time scale $ds=n\sigma ^{2}g v_{T}(t)dt$ has been introduced. Here,
 $\Phi \left( \beta ,h;\phi \right)$ is a {\em variational\/} free energy given
by Eq.\ (\ref{9}) with the order parameter $\phi$ considered as an independent variable,
and the kinetic coefficient $\mu (\phi )$ is 
\begin{equation}
\mu (\phi )=\nu^*{\left( 1+\phi \right) ^{3/2}}. 
\label{16}
\end{equation}
The stationary solution occurs for $\partial \Phi \left( \beta ,h;\phi \right)
/\partial \phi =0$, which is just Eq.\ (\ref{5}). For states near the
HCS the equation can be linearized and a characteristic response time $\tau$
identified according to 
\begin{equation}
-\partial _{s}\ln (\phi -\phi _{\text{HCS}})=\tau ^{-1}=\left( \mu \chi _{\beta
}^{-1}\right) _{\phi _{\text{HCS}}},  \label{17}
\end{equation}
where $\phi _{\text{HCS}}$ denotes the solution to (\ref{5}). In the elastic limit $
\tau $ is just the equilibration time for the impurity particle to attain a
mean kinetic energy equal to that of the fluid particles. Similarly, for
inelastic collisions it is the time (in units of the fluid mean free time) for the impurity particle to reach a
cooling rate equal to that of the fluid. This characteristic time is a
smooth function of $h$ and $\beta $ except in the limit $h\rightarrow 0$
where $\tau $ diverges at $\beta =1$. This critical slowing follows directly
from the fact that $\tau \propto \chi_\beta $. Otherwise, the relaxation times
away from $\beta =1$ are finite and comparable for the normal and ordered
states. Figure \ref{fig3} shows the dependence of the inverse time ${\tau^*}^{-1}\equiv (\nu^*
\tau)^{-1}$ on $h$ for $\beta=0.9$, 1, and 1.1. The quantity
$\tau^*$ measures the relaxation time for the impurity particle in terms of
 the number of impurity--fluid particle collisions.

\begin{figure}[t]
\centerline{\epsfig{file=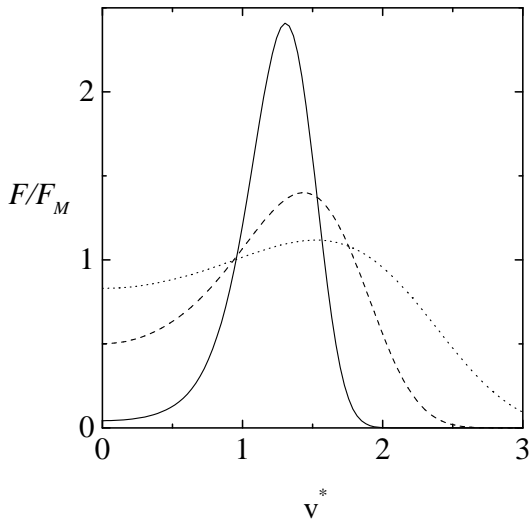,width=7.0cm}}
\caption{Velocity distribution function of the impurity particle, $F$, relative
to the Maxwellian, $F_M$, for
$h=10^{-2}$ and $\beta=0.9$ (dotted line), $\beta=1$ (dashed line), 
and $\beta=1.1$ (solid line).
\label{fig4}}
\end{figure}

The above analysis is based on an approximate evaluation of the cooling
rates using Maxwellians for the impurity and fluid particle velocity
distributions. This is known to be accurate for the fluid particles\cite{Brey1} but the
new features in the critical domain and in the ordered phase $\beta >1$
require closer inspection of the impurity particle distribution. Based on
the Lorentz-Enskog kinetic equation, it is possible to determine the
impurity particle velocity distribution for the HCS asymptotically close to $\beta
\rightarrow 1$ and $h\rightarrow 0$ with the result 
\begin{equation}
F({\bf v}^*)\rightarrow C\exp \left\{-\frac{\phi }{h}\left[ \left( 1-\beta \right)
{v^*}^2+
\frac{1}{10}\phi {v^*}^{4}\right]\right\},
\label{ 18}
\end{equation}
where ${\bf v}^*={\bf v}_0/v_{T0}$ and $C$ is a normalization constant. The asymptotic equation of state is
determined from $\int d{\bf v}^* \left({v^*}^2-\frac{3}{2}\right)F({\bf
v}^*)=0$.
The equation of state is similar to that of Eq.\ (\ref{6}) with only  minor
quantitative changes in some coefficients. This justifies the preceding
analysis based on the Maxwellian approximation to $\xi _{0}^{\ast }$.
Nevertheless, the distribution function itself shows new qualitative
differences. As expected, for $\beta <1,$ $\phi \rightarrow h/\left( 1-\beta
\right) $ and the Maxwellian is recovered. For $\beta >1$, $\phi \rightarrow
10\left( \beta -1\right) /3$ and the velocity distribution becomes sharp at $
v^*=\sqrt{3/2}$,
$F({\bf v}^*)\rightarrow (6\pi)^{-1}\delta
\left(v^*-\sqrt{{3}/{2}}\right)$.
Furthermore, at the critical point ($\beta=1$), $\phi\to \sqrt{10\lambda h}$ and
$F({\bf v}^*))\rightarrow C\exp (-\lambda {v^*}^{4})$,
where $\lambda \simeq 0.24$. This crossover from Gaussian to exponential
quartic to delta distributions is illustrated in Fig.\  \ref{fig4} for $h=10^{-2}$
and $\beta =0.9$, 1.0, and 1.1.

The origin of this transition can be traced to the fact that the cooling of the
fluid is due to collisions that depend on the restitution coefficient $%
\alpha $ but not the mass ratio (or $h$), while cooling of the impurity
depends on the mass ratio but not $\alpha $. The equality of cooling rates
is enforced by adjusting the ratio of the mean square velocities, $\phi $.
In general, for fixed mechanical properties, this is possible for small mass
ratio only if $\phi $ is non zero (the ordered state); the cooling rate for
the impurity has an explicit proportionality to $h$ which must be countered
by $\phi $. This implies that the mean kinetic energy of the impurity
particle diverges relative to that of the fluid particles. However, if this
explicit decrease of the impurity cooling rate with $h$ is tempered by a
large size ratio and/or weak fluid particle dissipation then the normal
behavior with $\phi =0$ is restored. The parameter $\beta $ quantifies the
meaning of ``tempered''. To give an explicit example, consider the case $%
\alpha =\alpha _{0}=0.95$ and $m/m_{0}=10^{-2}$. Also, for simplicity
consider low density so that $g=g_{0}\rightarrow 1$. Then $\beta =1.79\left(
\sigma /\overline{\sigma }\right) ^{2}$ and the system is close to the
ordered state for $\sigma =\sigma _{0}$ but close to the normal state for $%
\sigma_0 =2\sigma$. It is possible to study the crossover through the
critical domain using Direct Monte Carlo simulation of the Boltzmann-Enskog 
and Lorentz-Enskog kinetic equations, and by molecular dynamics
simulation. In fact both Monte Carlo molecular dynamics simulations for
the domain $\beta <1$ and $m/m_{0}\approx 10^{-2}$ have been performed to
confirm Brownian motion in the normal state \cite{Brey4}. Extension of these
studies to the critical and ordered states should be straightforward.

For experimental purposes it is perhaps more relevant to study these critical
properties as reflected in the diffusion coefficient $D$ (or mean square
displacement) for the impurity
particle. The dependence of $D$ on $\beta $ and $h$ can be calculated from
the Lorentz-Enskog kinetic equation in the same approximation as the
cooling rates above. The studies of the Fokker-Planck limit $\left(h
\rightarrow 0\right) $ in \cite{Brey3} suggest a divergence of $D$ as $\beta
\to 1$ although that analysis is strictly limited to $\beta <1$. A more
complete description of $D$ for all $\beta $, $h$  will be given elsewhere \cite{Santos}.

This research was supported by National Science Foundation grant PHY 9722133.
A.S. acknowledges partial support from the DGES (Spain) through grant No.\
PB97-1501
and through  a sabbatical grant
No.\ PR2000-0117.


\begin{references}
\bibitem{note2}Following the thermodynamic analogy, the critical exponents
$\hat{\delta}$, $\hat{\beta}$, $\hat{\gamma}$, and $\hat{\alpha}$ are
defined as $\phi(\beta=1,h)\sim h^{1/\hat{\delta}}$, $\phi(\beta>1,h=0)\sim
(\beta-1)^{\hat{\beta}}$, $\chi_\beta(\beta,h=0)\sim
|\beta-1|^{-\hat{\gamma}}$,
$C_h(\beta,h=0)\sim |\beta-1|^{-\hat{\alpha}}$.
\end{references}
\end{document}